# Analytic *l*-state solutions of the Klein–Gordon equation for q-deformed Woods-Saxon plus generalized ring shape potential

M. Chabab, A. Lahbas, M. Oulne[*]

High Energy Physics and Astrophysics Laboratory, Department of Physics, Faculty of Sciences Semlalia, Cadi Ayyad University, P.O.B 2390, Marrakesh 40000, Morocco

**Abstract:**

The analytical expressions for the eigenvalues and eigenvectors of the Klein-Gordon equation for q-deformed Woods-Saxon plus new generalized ring shape potential are derived within the asymptotic iteration method. The obtained eigenvalues are given in a closed form and the corresponding normalized eigenvectors, for any *l*, are formulated in terms of the generalized Jacobi polynomials for the radial part of the Klein-Gordon equation and associated Legendre polynomials for its angular one. When the shape deformation is canceled, we recover the same solutions previously obtained by the Nikiforov-Uvarov method for the standard spherical Woods-Saxon potential. It is also shown that, from the obtained results, we can derive the solutions of this problem for Hulthen potential.

Keywords: Klein-Gordon equation, q-deformed Woods-saxon potential, ring-shaped potential, asymptotic iteration method.

## I. Introduction.

The Woods–Saxon potential [1] is considered as the most realistic short range potential in nuclear physics. It is widely used in the study of the nuclear structure within the shell model. Besides, several generalized versions of this potential have been developed to explore elastic and quasi-elastic scattering of nuclear particles [2]. So, the woods–Saxon potential, either in its spherical or deformed form, has been more used in nuclear numerical calculations [3-7]. It was also applied in others domains of physics such as the study of the behavior of valence electrons in metallic systems or in helium model [8]. Moreover, it has been used in nonlinear scalar theory of mesons [9]. Recently, it has received increasing interest, in its central and spherical form, in the context of the conventional Nikiforov-Uvarov method, to derive analytical solutions of Schrödinger equation [10-13], Klein-Gordon equation [14-16] and Dirac equation [17-18]. Such a problem has also been solved by other methods [19-21].

In recent years, a new ring-shaped potential has been introduced [22] paving the way to new calculations in which this potential was combined with the Coulomb potential [22], Hulthen potential [23] or Kratzer potential [24]. Those calculations with this new non-central potential may have possible applications in quantum chemistry such as the study of ring-shaped molecules like benzene.

In nuclear physics, the shape form of the potential also plays an important role particularly when studying the structure of deformed nuclei or the interaction between them. Therefore, our aim, in the present work, is to investigate analytical bound state solutions of the Klein-Gordon (KG) equation with q-deformed Woods-Saxon potential related to a generalized ring-shaped potential, that we have obtained by a modification of the ring-shaped potential reported in [22], using the asymptotic iteration method (AIM) [25]. Also, we will show that, when the deformation parameter q takes a particular value, the obtained results lead to the solutions of the same problem for Hulthen potential.

The AIM, an increasingly popular method, has proven to be a powerful, efficient and easily handling method in the treatment of problems dealing with second order differential equations of type $y''(x) + k(x)y(x) = 0$ which are usually encountered in physics such as Schrödinger equation, the radial and angular parts of KG and Dirac equations [26-41].

This paper is organised as follows: in section 2, we present the KG equation for the q-deformed Woods–Saxon potential plus the generalized ring shaped potential. Section 3 is devoted to derive the exact analytic bound state energies and the corresponding radial solution of KG equation in the context of AIM method. In section 4, we solve the angular part of KG equation with relevant discussions. In section 5, we conclude our work.

[*]Corresponding author
E-mail addresses: oulne@ucam.ac.ma; mchabab@ucam.ac.ma; 3laaeddine@gmail.com

## II. The Klein–Gordon equation with a q-deformed Woods-Saxon plus a generalized ring shape potential

The Klein–Gordon equation of a particle with mass moving in a mixed scalar $S(r)$ and vector $V(r)$ potentials is given by:

$$\nabla^2 \psi(r) + \frac{1}{\hbar^2 c^2}\left\{(E - V(r))^2 - (m_0 c^2 + S(r))^2\right\}\psi(r) = 0 \quad (1)$$

In spherical coordinates, this equation reads

$$\left[\frac{1}{r^2}\frac{\partial}{\partial r}\left(r^2 \frac{\partial}{\partial r}\right) + \frac{1}{r^2 \sin\theta}\frac{\partial}{\partial \theta}\left(\sin\theta \frac{\partial}{\partial \theta}\right) + \frac{1}{r^2 \sin^2\theta}\frac{\partial^2}{\partial \varphi^2} + \right.$$
$$\left. \frac{1}{\hbar^2 c^2}\left[(E - V(r,\theta))^2 - (m_0 c^2 + S(r,\theta))^2\right]\right]\psi(r,\theta,\varphi) = 0 \quad (2)$$

where $E$ is the energy and $m_0$ the rest mass of the particle, c is the velocity of light and $\hbar$ the reduced Planck's constant.

The vector potential $V$ is chosen to be equal to the q-deformed Woods–Saxon potential ($V_{WS}$) plus the generalized ring-shaped one ($V_{RS}$), ($V = V_{WS} + V_{RS}$) while the scalar potential $S(r)$ is taken equal to $V_{RS}$. If $V_{RS}$ is considered as a small perturbation, the term $2V_{WS}V_{RS}$, appearing in the development of Eq.(2), can be neglected as a first approximation. Thus, the variables $r$, $\theta$ and $\varphi$ in Eq.(2) will be separable.

The generalized ring-shaped potential used in this work is obtained from the ring-shaped potential, established in [22] by adding some parameter α. Thus, the q-deformed Woods–Saxon potential [1] plus the generalized ring-shaped one is expressed as

$$V(r,\theta) = -\frac{V_0}{1 + q e^{\frac{r-R_0}{a}}} + \frac{\alpha + \beta \cos^2\theta}{r^2 \sin^2\theta} \quad (3)$$

where the first term in the right hand side of Eq.(3) is the q-deformed Woods–Saxon potential with $V_0$ is the potential depth, $R_0$ represents the width of the potential or the nuclear radius, a surface thickness parameter and the parameter q defines the deformation of the potential. The second term introduces the short range ring shaped potential [22] modified with an extra additional parameter α.

In order to separate the variables in Eq.(2), the particle wave function can be chosen in the following form

$$\psi(r,\theta,\varphi) = \frac{R(r)}{r} Y(\theta,\varphi)$$

where the angular part of this function is selected in the form $Y(\theta,\varphi) = \Theta(\theta)\Phi(\varphi)$.

Then, the radial part of Eq.(2) is given by,

$$\left[\frac{d^2}{dr^2} + \frac{(E - V_{WS})^2 - m_0^2 c^4}{\hbar^2 c^2} - \frac{\ell(\ell+1)}{r^2}\right]R(r) = 0 \quad (5)$$

while the angular equations read, assuming $2V_{RS} \approx V_{RS}$, as

$$\left[\frac{d^2}{d\theta^2} + \cot\theta \frac{d}{d\theta} + \ell(\ell+1) - \frac{m^2}{\sin^2\theta} - (E + m_0 c^2)\frac{\alpha + \beta \cos^2\theta}{\sin^2\theta}\right]\Theta(\theta) = 0 \quad (6.a)$$

$$\left[\frac{d^2}{d\varphi^2} + m^2\right]\Phi(\varphi) = 0 \quad (6.b)$$

where $\ell$ is the angular momentum quantum number and m a constant.

## III Solution of the radial equation

To solve the radial equation (Eq.(5)), we introduce the following conversions

$$x = \frac{r - R_0}{R_0}, \quad \alpha = \frac{R_0}{a} \quad (7)$$

Thus the Woods – Saxon potential in Eq.(3) transforms into,

$$V(x) = -\frac{V_0}{1+qe^{\alpha x}} \tag{8}$$

In order to simplify the solution of Eq.(5), the centrifugal potential will be developed in powers of $y = \frac{1}{1+qe^{\alpha x}}$ by means of the Pekeris approximation [42-43]

$$V_\ell(x) = \frac{\hbar^2 \ell(\ell+1)}{2mR_0^2(1+x)^2} = \frac{\hbar^2 \ell(\ell+1)}{2mR_0^2}(C_0 + C_1 y + C_2 y^2) \tag{9}$$

where the conversions (7) have been used. The coefficients $C_0$, $C_1$ and $C_2$ are determined through the comparison of the series developments of the right – hand and the left – hand sides of the Eq.(8) around x = 0. We get

$$\begin{aligned}
C_0 &= 1 - \left(3 + \frac{2}{q} - \frac{1}{q^2}\right)\frac{1}{\alpha} + 3\left(1 + \frac{2}{q} + \frac{1}{q^2}\right)\frac{1}{\alpha^2}, \\
C_1 &= 2\left(3 + 2q - \frac{1}{q^2}\right)\frac{1}{\alpha} - 6\left(3 + q + \frac{3}{q} + \frac{1}{q^2}\right)\frac{1}{\alpha^2}, \\
C_2 &= -\left(2q + 3q^2 - \frac{2}{q} + \frac{1}{q^2}\right)\frac{1}{\alpha} + 3\left(6 + 4q + q^2 + \frac{4}{q} + \frac{1}{q^2}\right)\frac{1}{\alpha^2}
\end{aligned} \tag{10}$$

Inserting Eq.(8) and Eq.(9) into Eq.(5) and using the new variable y, the Eq.(5) becomes

$$y(1-y)\frac{d}{dy}\left(y(1-y)\frac{dR(y)}{dy}\right) - \frac{\varepsilon^2 + \beta^2 y + \gamma^2 y^2}{y(1-y)}R(y) = 0 \tag{11}$$

where the following transformations have been done

$$\varepsilon^2 = -\frac{(E^2 - m_0^2 c^4)a^2}{\hbar^2 c^2} + \frac{\ell(\ell+1)C_0}{\alpha^2}, \beta^2 = -\frac{2EV_0 a^2}{\hbar^2 c^2} + \frac{\ell(\ell+1)C_1}{\alpha^2}, \gamma^2 = -\frac{V_0^2 a^2}{\hbar^2 c^2} + \frac{\ell(\ell+1)C_2}{\alpha^2} \tag{12}$$

The boundary conditions for the radial wave function are $R(0) = 0$ and $R(\infty) = 0$.
To solve the Eq.(11) by means of the asymptotic iteration method [25], we propose the following ansatz for $R(y)$

$$R(y) = Ay^\mu(1-y)^\sigma f(y) \tag{13}$$

where $A$ is a normalization constant and $\mu$ and $\sigma$ are some parameters to be defined latter.
For this form of the wave function, the radial equation (11) reads

$$\frac{d^2}{dy^2}f(y) = \lambda_0(y)\frac{d}{dy}f(y) + s_0(y)f(y) \tag{14}$$

with

$$\lambda_0(y) = -\frac{1 + 2\mu - 2(\mu + \sigma + 1)y}{y(1-y)} \tag{15a}$$

$$s_0(y) = \frac{(\mu+\sigma)(\mu+\sigma+1) - v^2}{y(1-y)} \tag{15b}$$

where

$$\mu = \varepsilon, \sigma = \sqrt{\varepsilon^2 + \beta^2 + \gamma^2}, v = \gamma \tag{16}$$

Then according to the AIM procedure, the energy eigenvalues are then calculated by means of the following termination condition [25]

$$\delta = s_n \lambda_{n-1} - \lambda_n s_{n-1} = 0 \tag{17}$$

for a given y > 1, with the sequences

$$\lambda_n(y) = \lambda'_{n-1}(y) + s_{n-1}(y) + \lambda_0(y)\lambda_{n-1}(y) \tag{18a}$$

$$s_n(y) = s'_{n-1}(y) + s_0(y)\lambda_{n-1}(y), \text{ n= 1, 2, 3, ...} \tag{18b}$$

For a few iterations, the obtained solutions are

$$\mu_0 = -\sigma_0 - \frac{1}{2} \pm \frac{1}{2}\sqrt{1+4v^2}, \mu_1 = -\sigma_1 - \frac{3}{2} \pm \frac{1}{2}\sqrt{1+4v^2}, \mu_2 = -\sigma_2 - \frac{5}{2} \pm \frac{1}{2}\sqrt{1+4v^2},... \tag{19}$$

from which the following general form is easily deduced

$$\mu_n = -\sigma_n - \frac{2n+1}{2} \pm \frac{1}{2}\sqrt{1+4v^2}, \ n = 0, 1, 2,..... \tag{20}$$

Substituting $\mu$, $\sigma$ and $v$ by their expressions given in Eq. (16), we obtain the eigenvalues equation

$$\varepsilon_n + \sqrt{\varepsilon_n^2 + \beta^2 + \gamma^2} = -\frac{2n+1}{2} \pm \frac{1}{2}\sqrt{1+4\gamma^2} = N \tag{21}$$

where n (n = 0, 1, 2, …) is the radial quantum number and

$$N = -n - \frac{1}{2} + \frac{1}{2}\sqrt{1+4\gamma^2} \tag{22}$$

The solution of Eq.(21) gives

$$\varepsilon_N = \frac{1}{2}\left(N - \frac{\beta^2+\gamma^2}{N}\right) \tag{23}$$

Once the expressions of $\varepsilon$, $\beta$ and $\gamma$, given in Eq. (12), are substituted into Eq. (23), we get the generalized formula of the radial energy eigenvalues,

$$E_{n,\ell} = -\frac{V_0}{2}\left(1 - \frac{4(C_1+C_2)a^2\ell(\ell+1)}{R_0^2\left(N'^2 + \frac{4V_0^2a^2}{\hbar^2c^2}\right)}\right) \mp \frac{N'}{N'^2 + \frac{4V_0^2a^2}{\hbar^2c^2}} \times$$

$$\left\{\left[\left[m_0^2c^4 + \frac{\hbar^2c^2\ell(\ell+1)\left(C_0+\frac{1}{2}C_1+\frac{1}{2}C_2\right)}{R_0^2}\right]\left(N'^2 + \frac{4V_0^2a^2}{\hbar^2c^2}\right)^{-1} - \frac{1}{16}\frac{\hbar^2c^2}{a^2}\right]\left(N'^2 + \frac{4V_0^2a^2}{\hbar^2c^2}\right)^2 - \left[\frac{\hbar c a \ell(\ell+1)(C_1+C_2)}{R_0^2}\right]^2\right\}^{1/2} \tag{24}$$

where $N' = 2N = \sqrt{1 - \frac{4V_0^2a^2}{\hbar^2c^2} + \frac{4\ell(\ell+1)C_2}{\alpha^2}} - (2n+1)$. In the standard case of the spherical Woods-Saxon potential where q = 1, our energy spectrum formula matches up with the results of Nikifrov-Uvarov approach [44] (with $\hbar=c=1$).

For $q = -e^{\alpha R_0}$ and setting $V_1 = -V_0$, the q-deformed Woods-Saxon potential (Eq.(3)) transforms into Hulthen potential [45]. For $\alpha R_0 \to 0$, the coefficients, in the series development of the centrifugal potential (Eq.(10)), tend to $C_0 = 1$ and $C_1 = C_2 = 0$ and from Eq.(24), we can get an identical energy eigenvalues formula for Hulthen potential to that obtained in [46-47] for s-wave ($\ell = 0$) taking $N' = 2N = \sqrt{1 - \frac{4V_0^2a^2}{\hbar^2c^2}} + (2n+1)$ and the Hulthen potential constant $\alpha = 1/a$.

The eigenvectors of the problem are derived through the determination of the expression of the AIM function $f(x)$ by solving the differential equation (14)

$$\frac{d^2}{dy^2}f(y) + \left(\frac{1+2\mu-2(\mu+\sigma+1)y}{y(1-y)}\right)\frac{d}{dx}f(y) - \left(\frac{(\mu+\sigma)(\mu+\sigma+1)-v^2}{y(1-y)}\right)f(y) = 0 \tag{25}$$

whose solutions, in the limit $x \to 0$ are the hyper-geometrical functions,

$$f(y) = C \ _2F_1(-n, 1+2\mu+2\sigma+n; 2\mu+1; x) \tag{26}$$

Therefore, according to the relation between hyper-geometrical functions and the generalized Jacobi polynomials $P_n^{(\alpha,\beta)}(z)$ [48-49], the radial wave function can be written as

$$R_{n\ell}(y) = A_{n\ell}y^\varepsilon(1-y)^{\sqrt{\varepsilon^2+\beta^2+\gamma^2}} P_n^{\left(2\varepsilon, 2\sqrt{\varepsilon^2+\beta^2+\gamma^2}\right)}(1-2y) \tag{27}$$

where $A_{n\ell}$ is a normalization constant computed via the orthogonality relation of Jacobi polynomials [48-49]

$$A_{n\ell} = \frac{(2\mu+n-1)!}{(2\mu-1)!} \left[\frac{n+\mu+\sigma+\frac{1}{2}}{n!} \frac{\Gamma(n+2(\mu+\sigma)+1)}{\Gamma(n+2\mu+1)\Gamma(n+2\sigma+1)}\right]^{1/2} \tag{28}$$

where $\mu$ and $\sigma$ are given in Eq.(16).

**IV Solution of the angular equation**

To calculate the eigenvalues of the angular equation (6a), we introduce a new variable $= \cos\theta$, so we obtain the following equation

$$\left[(1-z^2)\frac{d}{dz}(1-z^2)\frac{d}{dz} + \upsilon(1-z^2) - \mu^2\right]H(z) = 0 \tag{29}$$

where

$$\mu = \sqrt{m^2 + \alpha' + \beta'} \ , \upsilon = \beta' + \ell(\ell+1) \tag{30}$$

with $\alpha' = (E + m_0 c^2)\alpha$ and $\beta' = (E + m_0 c^2)\beta$ and $H(z)$ undergoes the boundary conditions $H(0) = H(1) = 0$. Moreover, to solve Eq.(29) with the asymptotic iteration method [25] we use the following ansatz

$$H(z) = B(1-z^2)^{\frac{\mu}{2}} f(z) \tag{31}$$

which leads after its substitution into Eq.(29) to

$$\frac{d^2}{dy^2}f(z) = \lambda_0(z)\frac{d}{dy}f(z) + s_0(z)f(z) \tag{32}$$

where

$$\lambda_0(z) = \frac{2(\mu+1)z}{(z+1)(1-z)} \tag{33a}$$

$$s_0(z) = \frac{\mu^2+\mu-\upsilon}{(z+1)(1-z)} \tag{33b}$$

The eigenvalues of Eq.(29) are determined by solving the AIM equation (Eq.(17)) for a given $z > 1$. For a few iterations we obtain

$$\upsilon = \mu^2 + \mu, \upsilon = \mu^2 + 3\mu + 2, \upsilon = \mu^2 + 5\mu + 6, \upsilon = \mu^2 + 7\mu + 12, \ldots \tag{34}$$

In general form we have

$$\upsilon = (\mu+n)(\mu+n+1); n = 0,1,2,3,\ldots \tag{35}$$

Replacing $\mu$ and $\upsilon$ by their expressions given in (30) we find

$$\ell_n = -\frac{1}{2} \pm \sqrt{\left(n + \sqrt{m^2 + \alpha' + \beta'} + \frac{1}{2}\right)^2 - \beta'} \tag{36}$$

Substituting the obtained expression for $\ell$ (Eq.(36)) into Eq.(24) we get the energy eigenvalues for a bound particle in the q-deformed Woods-Saxon plus generalized ring-shape potential.

$$E_{N',n,m} = -\frac{V_0}{2}\left(1 - \frac{4(C_1+C_2)a^2\left(\left(n+\sqrt{m^2+\alpha'+\beta'}+\frac{1}{2}\right)^2 - \beta' - \frac{1}{4}\right)}{R_0^2\left(N'^2 + \frac{4V_0^2 a^2}{\hbar^2 c^2}\right)}\right) \mp \frac{N'}{N'^2 + \frac{4V_0^2 a^2}{\hbar^2 c^2}} \times$$

$$\left\{\left[\left[m_0^2 c^4 + \frac{\hbar^2 c^2\left(\left(n+\sqrt{m^2+\alpha'+\beta'}+\frac{1}{2}\right)^2 - \beta' - \frac{1}{4}\right)\left(C_0+\frac{1}{2}C_1+\frac{1}{2}C_2\right)}{R_0^2}\right]\left(N'^2 + \frac{4V_0^2 a^2}{\hbar^2 c^2}\right)^{-1} - \frac{1}{16}\frac{\hbar^2 c^2}{a^2}\right]\left(N'^2 + \frac{4V_0^2 a^2}{\hbar^2 c^2}\right)^2 - \left[\frac{\hbar c a\left(\left(n+\sqrt{m^2+\alpha'+\beta'}+\frac{1}{2}\right)^2 - \beta' - \frac{1}{4}\right)(C_1+C_2)}{R_0^2}\right]^2\right\}^{1/2} \qquad (37)$$

The eigenvectors of the problem are determined by solving the following associated Legendre differential equation which is obtained from Eq.(29) [48-49]

$$(1-z^2)\frac{d^2}{dz^2}H(z) - 2z\frac{d}{dz}G(z) + \left[\ell'(\ell'+1) - \frac{\mu^2}{1-z^2}\right]H(z) = 0 \qquad (38)$$

where $\ell' = \frac{1}{2}\sqrt{4(\ell(\ell+1) - \beta') + 1} - \frac{1}{2}$ or through the solution of the Eq.(32) and using the ansatz proposed in Eq.(31). The both ways yield to the following associated Legendre polynomials [48-49]

$$H(z) = \frac{(-1)^\mu}{2^{\ell'}\ell'!}(1-z^2)^{\mu/2}\frac{d^{\ell'+\mu}}{dz^{\ell'+\mu}}(z^2-1)^{\ell'} \qquad (39)$$

where $\mu$ and $\upsilon$ are given in Eq.(30).

**VI Conclusion**

In this paper we have solved the Klein-Gordon equation with a q-deformed Woods-Saxon potential plus a new generalized ring-shaped potential. The energy eigenvalues of this problem are obtained in analytic formulas and the corresponding radial eigenvectors are formulated in terms of generalized Jacobi polynomials while the angular eigenvectors are expressed in closed form through associated Legendre polynomials. When the shape deformation is canceled (q = 1), the energy eigenvalues are identical to those obtained by the Nikiforov-Uvarov method. Also, when the deformation parameter is set to the particular value $q = -e^{\alpha R_0}$, we have shown that we can deduce the analytical solutions corresponding to Hulthen potential . The obtained solutions of the present problem may find applications in many areas of physics particularly in the study of the interaction between deformed nuclear particles.

.